\documentclass{article}
\usepackage{ijcai26}

\usepackage{times}
\usepackage{soul}
\usepackage{url}
\usepackage[hidelinks]{hyperref}
\usepackage[utf8]{inputenc}
\usepackage[small]{caption}
\usepackage{graphicx}
\usepackage{amsmath}
\usepackage{amsthm}
\usepackage{booktabs}
\usepackage{algorithm}
\usepackage{algorithmic}
\usepackage[switch]{lineno}

\usepackage{multirow}
\usepackage{endnotes}
\usepackage{float}
\usepackage[accsupp]{axessibility}
\usepackage{enumitem}
\usepackage{xcolor}
\definecolor{customgreen}{rgb}{0.32,0.74,0.32}

\title{UniField: Joint Multi-Domain Training for Universal Surface Pressure Modeling}

\author{
Junhong Zou$^{1,2}$\and
Zhenxu Sun$^3$\and
Yueqing Wang$^4$\and
Wei Qiu$^3$\and
Zhaoxiang Zhang$^{1,2}$\and\\
Xiangyu Zhu$^{1,2*}$\And
Zhen Lei$^{1,2}$\footnote{Corresponding authors}\\
\affiliations
$^1$MAIS, Institute of Automation, Chinese Academy of Sciences, Beijing, China,\\
$^2$School of Artificial Intelligence, University of Chinese Academy of Sciences, Beijing, China,\\
$^3$Key Laboratory for Mechanics in Fluid Solid Coupling Systems, Institute of Mechanics, Chinese\\Academy of Sciences, Beijing, China,\\
$^4$State Key Laboratory of Aerodynamics, Sichuan, China\\
\emails
\{zoujunhong2022, xiangyu.zhu, zhen.lei\}@ia.ac.cn
}

\begin{document}

\maketitle

\begin{abstract}
	Accurate modeling of surface pressure fields around objects is fundamental to aerodynamic analysis and design.
	While neural networks have shown promise as efficient alternatives to expensive Computational Fluid Dynamics (CFD) simulations,
	their applicability is often constrained by data scarcity and poor generalization across different aerodynamic domains.
	To address these challenges, we propose UniField, a unified framework that enables joint training across multiple aerodynamic domains including automobiles, trains, aircraft.
	UniField employs a shared geometry encoder to extract domain-agnostic representations from surface point clouds,
	and integrates domain-specific flow information through Parallel Flow-Conditioned Adaptive LayerNorm (PFC-AdaLN).
	In addition to consolidating existing datasets from specialized research field including automobiles, trains and aircraft, we further introduce ThingiCFD, a large-scale CFD dataset constructed from Thingi10k geometries with extensive flow condition randomization,
	substantially expanding geometric and flow diversity during training.
	UniField achieves SOTA performance on the public DrivAerNet++ benchmark. In addition, our experiments demonstrate that joint multi-domain training consistently improves surface pressure prediction accuracy,
	particularly in data-scarce domains.
	These results highlight the potential of UniField as a foundation model for data-driven aerodynamic modeling. Code and data will be available at \href{https://github.com/zoujunhong/UniField}{https://github.com/zoujunhong/UniField}.
\end{abstract}

\begin{figure}[t!]
	\centering
	\includegraphics[width=1 \columnwidth]{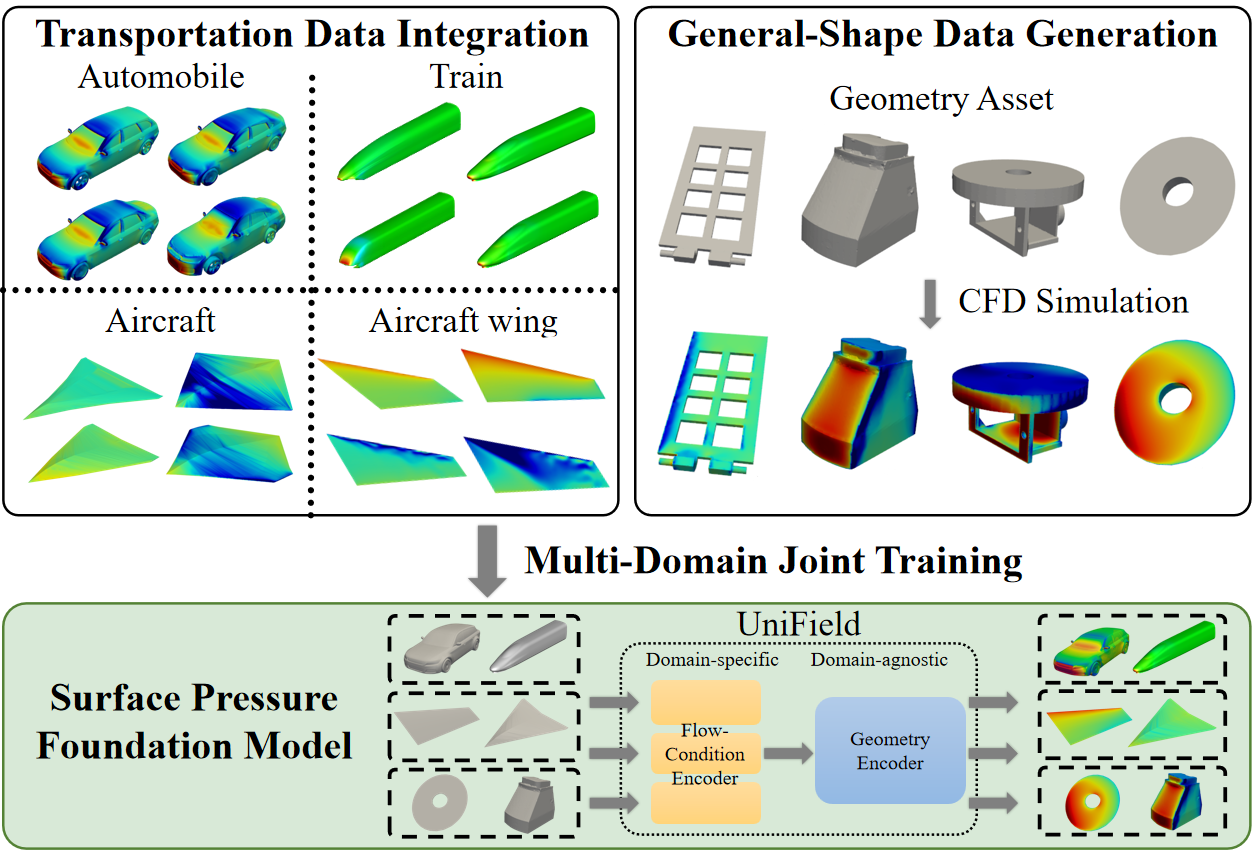}
	\caption{Overview. UniField is a surface pressure foundation model pre-trained on multiple aerodynamic datasets covering diverse geometry and flow regimes (e.g., automobiles, trains, aircraft, and general-shape objects). The model captures unified representation of geometry and flow field, exhibiting improved generalization and better surface pressure predictions across unseen geometries under data scarcity situations.}
	\label{fig: summary}
\end{figure}    
\section{Introduction}
\label{introduction}

Modeling surface pressure fields around complex objects plays a central role in aerodynamic analysis,
with direct implications for performance evaluation and design optimization in applications such as automobiles, trains, and aircraft \cite{qiu2025compressed,gao2025accurate}.
High-fidelity Computational Fluid Dynamics (CFD) simulations remain the standard approach for obtaining accurate pressure distributions,
but their high computational cost severely limits scalability across diverse geometric configurations and flow conditions.

Recent advances in deep learning \cite{TripNet,FIGConvNet,Wang_Zhang_Liu_Zhao_Lin_Chen_2025,DrivAerNet++} have demonstrated that neural networks can approximate aerodynamic fields orders of magnitude faster than traditional CFD solvers.
However, most existing data-driven approaches are trained on narrowly scoped datasets and struggle to generalize to novel geometries or flow conditions~\cite{zhang2024prediction,qiu2025compressed,ZHANG2025100571,10.1063/5.0091980}.
This limitation stems from the inherent siloing of aerodynamic data across application domains. Beyond incompatible flow conditions, a more fundamental cause lies in the drastic variation of characteristic velocity scales across domains (e.g., automobiles $\ll$ high-speed trains $\ll$ aircraft), leading to qualitatively different flow regimes and governing assumptions, such as incompressible modeling for ground vehicles versus compressible flow for aircraft.
This situation results in a domain-specific research paradigm where models are developed exclusively to solve problems within a single field.


On the other hand, recent advances in deep learning suggest that increasing model capacity and training on diverse data can lead to the emergence of strong generalization and even domain-agnostic representations, raising the possibility of unifying knowledge across traditionally separated aerodynamic domains within a single neural model.
However, achieving such unification remains highly challenging in practice, as aerodynamic datasets across different domains are governed by different flow regimes and physical assumptions. For instance, the flow around an automobile is typically modeled as incompressible and described by velocity, whereas aircraft flows involve compressible regimes characterized by Mach number and Angle of Attack. The aforementioned data siloing issue also causes researchers to focus only on the data within their own fields, lacking multi-field joint datasets.
These discrepancies make it nontrivial for a single network to jointly learn from multi-domain data.

To effectively utilize aerodynamic data from multiple fields, we propose \textbf{UniField}, a unified framework that enables learning through integrated multi-domain aerodynamic data.
At its core, UniField adopts a domain-agnostic geometric encoder that operates directly on surface point clouds and leverages a window-based attention mechanism, enabling efficient and high-throughput processing of large-scale point clouds.
To accommodate the substantial differences in flow conditions across different domains, UniField incorporates Parallel Flow-Conditioned Adaptive LayerNorm (PFC-AdaLN) modules, which modulate the geometric features using domain-specific flow information in a modular manner.
By combining a domain-agnostic geometric backbone with domain-specific flow condition injection, UniField supports stable joint training across diverse aerodynamic domains while preserving a unified latent representation for effective cross-domain knowledge collaboration.

As shown in Figure \ref{fig: summary}, we integrate aerodynamic datasets from multiple application domains for training UniField, including automobiles~\cite{DrivAerNet++}, trains~\cite{qiu2025compressed}, and aircraft~\cite{Wang_Zhang_Liu_Zhao_Lin_Chen_2025}.
Still, existing surface pressure datasets are predominantly drawn from specialized engineering domains, where geometric variations are largely confined to a limited set of canonical configurations, resulting in insufficient geometric diversity.
To further enrich the training data, we propose \textbf{ThingiCFD}, a large-scale CFD dataset constructed from the diverse object geometries in the real-world 3D printing dataset Thingi10k~\cite{Thingi10K}.
ThingiCFD expands geometric coverage by incorporating generic object shapes and generating corresponding surface pressure data through CFD simulations. By incorporating randomized aerodynamic parameter selection and geometric transformations during the simulation process, ThingiCFD enables models to be exposed to a wider range of aerodynamic scenarios during training.

After joint training on multiple datasets, UniField demonstrates the ability to learn universal flow field representations and enhance the performance of each subfield. On the one hand, UniField achieves SOTA performance when compared with other existing models on the public benchmark DrivAerNet++. On the other hand, we compare the performance of UniField trained with the joint dataset against the case trained with data from a single domain. Keeping other settings constant, the joint-trained model exhibits superior performance across all subfields, and this advantage is more pronounced in data-scarce domains. For instance, in the \textit{Train} dataset~\cite{qiu2025compressed} the joint-trained model reduced the error by more than 50\% on the test set compared to the model trained only with train data. 

To summarize, we make the following contributions:
\begin{itemize}
	\item We propose \textbf{UniField}, a universal framework for fluid field analysis around objects, which enables cross-domain joint training via a domain-agnostic geometry feature extractor coupled with domain-specific flow condition encoders.
	\item We construct and integrate a diverse collection of aerodynamic datasets spanning vehicles, trains and aircrafts, and propose \textbf{ThingiCFD}, a large-scale general-shape CFD dataset, supporting joint multi-domain training.
	\item Extensive experiments demonstrate that joint multi-domain training consistently improves surface pressure prediction accuracy, with particularly pronounced gains in data-scarce domains such as train and aircraft flows, highlighting the effectiveness of multi-domain data integration for alleviating CFD data scarcity.
\end{itemize}

\begin{figure*}[t!]
	\centering
	\includegraphics[width=2.1\columnwidth]{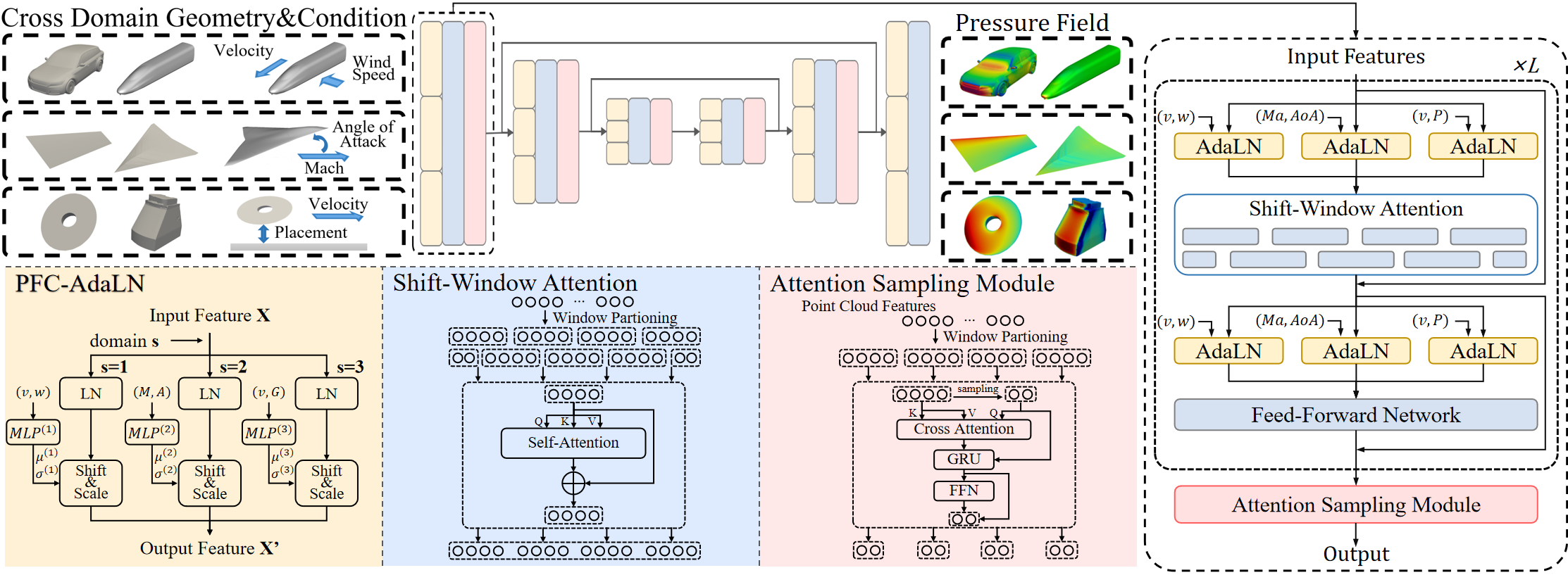}
	\caption{UniField Architecture. UniField learns with multi-domain data, including automobiles, aircraft, and general-shape objects. Given point cloud as input, UniField uses a UNet-style network to predict dense pressure fields. Each layer extracts the geometric features with shift-window attention. PFC-AdaLN are inserted to inject domain-specific information, thereby achieving cross-domain adaptation. The attention-based sampling module iteratively interacts between semantically related points achieve downsampling or upsampling.}
	\label{fig: architecture}
\end{figure*}

\section{Related Work}

\subsection{Surface Pressure Prediction}
Surface pressure prediction aims to infer the pressure distribution across an object’s surface given its geometry and flow conditions. Traditional CFD software simulates surface fields by numerically solving the Navier–Stokes equations \cite{constantin1988navier} over a discretized computational domain, but is computationally heavy, motivating neural networks as alternatives. DrivAerNet++ \cite{DrivAerNet++} is a large-scale aerodynamic dataset. It provides thousands of car geometries, CFD flow, pressure fields, and aerodynamic coefficients, fostering robust training and evaluation for automotive aerodynamics. It also introduce RegDCGNN \cite{Elrefaie_2025}, a dynamic graph convolutional neural network to regress aerodynamic parameters.
In the subsequent work, Transolver \cite{Transolver,luo2025transolveraccurateneuralsolver} leverages a Transformer-based PDE solver with a physics-inspired slice attention that groups mesh points into learnable physical-state slices. 
Factorized Implicit Global Convolution (FIGConvNet) \cite{FIGConvNet} efficiently learns global interactions across 3D meshes via implicit factorization, reducing complexity while preserving accuracy.
TripNet \cite{TripNet} encodes 3D car geometry into compact triplane representations, enabling point-wise predictions of pressure and full flow fields.

\subsection{Generalization of Neural Models for Flow Field Modeling}

With the development of research, deep learning methods have been proposed for addressing various aerodynamic tasks, while their applicability and generalization ability are gradually expanding. We summarize these models into the following categories. 

\begin{itemize}
	\item \textbf{Models for Single-Scene or Fixed Geometry.} Some models are only designed for Single-Scene or Fixed Geometry. For instance, PROSNet~\cite{zou2025joint} optimizes the reconstruction of the field and the arrangement of sensors for a specific high-speed train geometry, and PINNs \cite{RAISSI2019686,Yan_2024} explicitly embeds partial differential equation (PDE) constraints into the loss function to enforce physical consistency. While effective for a single geometry or under fixed boundary conditions, their model parameters are tightly coupled with specific physical settings.
	
	\item \textbf{Neural Operators.} Neural operator approaches~\cite{li2020fourier,wang2024latent,tran2021factorized} aim to learn the mapping between input and output fields~\cite{kovachki2023neural}, enabling fast approximations of specific PDEs. They offer broader applicability, but remain constrained to fixed governing equations, making them difficult to transfer across fundamentally different physical regimes (e.g., incompressible vs. compressible flows, subsonic vs. transonic regimes).
	
	\item \textbf{Symbolic or Parameterized PDE-Driven Solvers.} Recent frameworks such as UniSolver~\cite{zhou2024unisolver} and PDEformer~\cite{ye2025pdeformer} attempt to introduce the symbolic or parametric representation of PDEs as model inputs, moving toward general PDE solvers. Such methods still rely on complete prior knowledge of the governing equations and precise boundary conditions.
	
	\item \textbf{Data-Driven Field Modeling.}
	A growing line of research \cite{TripNet,FIGConvNet,Wang_Zhang_Liu_Zhao_Lin_Chen_2025} seeks to eliminate explicit dependence on predefined PDE by reformulating flow-field prediction as a regression problem. These models directly take geometries as inputs to predict target fields such as pressure and velocity. This formulation opens the possibility for building foundation models, yet research in this direction remains limited. Our proposed UniField follows this paradigm, leveraging multi-domain data and joint training to achieve cross-domain generalization.
\end{itemize}

\section{Method}

\subsection{Problem Setup}

Each input sample consists of a surface point cloud
$\mathcal{P}=\{\mathbf{p}_i \in \mathbb{R}^3\}_{i=1}^N$
and a domain-specific flow condition vector
$\mathbf{c}^{(s)} \in \mathbb{R}^{D_s}$,
such as $(v,w)$ (\textbf{v}elocity and \textbf{w}ind speed) for trains, $(Ma, AoA)$ (\textbf{Ma}ch number and \textbf{A}ngle \textbf{o}f \textbf{A}ttack) for aircraft, or $(v,P)$ (\textbf{v}elocity and \textbf{P}lacement) for general-shape data. In addition, a domain identifier $s \in \{1,\dots,M\}$, indicating one of $M$ aerodynamic subfields is also provided.
The model needs to predict a scalar pressure value for each point,
$\hat{\mathbf{y}} \in \mathbb{R}^N$. UniField is trained to minimize an $\ell_1$ regression loss between the predicted and ground-truth surface pressure fields.

\subsection{Architecture Overview}

As shown in Figure~\ref{fig: architecture}, we present UniField, a unified point-cloud-based framework for surface pressure field prediction across multi aerodynamic domains. The overall architecture of UniField follows a UNet-style hierarchical design, enabling dense point-wise prediction while maintaining scalability to large point clouds. It adopts a domain-agnostic geometry encoder with window-based attention and integrates domain-specific flow information through Parallel Flow-Conditioned Adaptive LayerNorm (PFC-AdaLN) before each attention and feed forward layers. We also propose an Attention Sampling Module for up- and down-sampling of the point clouds, enabling better information integration.

\subsection{Window-Based Geometry Encoder}

Given an input surface point cloud $\mathcal{P}=\{p_i\in\mathbb{R}^3\}_{i=1}^N$, we first embed each point coordinate into a $D$-dimensional feature space via a linear projection:
\begin{equation}
	\mathbf{x}_i^{(0)} = \mathbf{W}_{\text{in}}\, p_i + \mathbf{b}_{\text{in}},
	\label{eq:input_proj}
\end{equation}
where $\mathbf{x}_i^{(0)}\in\mathbb{R}^D$ denotes the initial feature of point $i$.

Recent point cloud processing methods \cite{chen2023pointgpt,wu2024point} propose treating point clouds as a 1D sequence and use window attention to extract features. Instead of their common practice of utilizing space-filling curves such as Morton (Z-order) ordering to serialize the point cloud, we randomly permute the points, which empirically yields better performance for surface pressure prediction, as it facilitates global geometric interaction, a property that is particularly important for aerodynamic pressure modeling compared to locality-preserving space-filling curves.

To facilitate efficient point cloud feature extraction, we process the resulting sequence with shift-window attention \cite{liu2021swin}. Specifically, the point sequence is partitioned into non-overlapping windows of size $P$, yielding $G = N/P$ windows. Within each window, UniField applies a Transformer block for feature extraction. For the $l$-th block, given input features $\mathbf{X}^{(l)}\in\mathbb{R}^{P\times D}$, the attention operation is defined as
\begin{equation}
	\mathrm{Attn}(\mathbf{Q},\mathbf{K},\mathbf{V}) = 
	\mathrm{Softmax}\!\left(\frac{\mathbf{Q}\mathbf{K}^\top}{\sqrt{D_h}}\right)\mathbf{V},
	\label{eq:window_attn}
\end{equation}
where $\mathbf{Q},\mathbf{K},\mathbf{V}$ are linear projections of $\mathbf{X}^{(l)}$, and $D_h$ is the per-head feature dimension. Each block follows a standard Transformer residual structure:

\begin{align}
	\tilde{\mathbf{X}}^{(l)} &= \mathbf{X}^{(l)} + 
	\mathrm{Attn}\!\left(\mathrm{Norm}(\mathbf{X}^{(l)})\right), \\
	\mathbf{X}^{(l+1)} &= \tilde{\mathbf{X}}^{(l)} +
	\mathrm{FFN}\!\left(\mathrm{Norm}(\tilde{\mathbf{X}}^{(l)})\right),
	\label{eq:window_block}
\end{align}
where $\mathrm{FFN}(\cdot)$ denotes a two-layer feed-forward network with GELU activation. Finally, to achieve the shifted-window strategy, consecutive blocks alternate between standard window partitioning and cyclically shifted windows, ensuring long-range interactions while maintaining linear complexity.

\subsection{Attention Sampling Module}

UniField employs an attention-based sampling module for both down-sampling and up-sampling of point features. Instead of relying on fixed geometric pooling rules, this module performs feature aggregation through cross-attention and recurrent updates.

Let $\mathbf{X}\in\mathbb{R}^{N\times D}$ denote the input point features, and let $\mathbf{S}^{(0)}\in\mathbb{R}^{M\times D}$ be a set of initial sampling features. The module iteratively refines $\mathbf{S}$ through cross-attention:
\begin{align}
	&\mathbf{Q} = \mathbf{S}^{(t)} \mathbf{W}_Q,\quad
	\mathbf{K} = \mathbf{X}\mathbf{W}_K,\quad
	\mathbf{V} = \mathbf{X}\mathbf{W}_V, \\
	&\mathbf{U}^{(t)} = 
	\mathrm{Attn}(\mathbf{Q},\mathbf{K},\mathbf{V}),
	\label{eq:cross_attn}
\end{align}
where $\mathbf{W}_Q,\mathbf{W}_K,\mathbf{W}_V$ are learnable projections.

The sampling features are then updated via a gated recurrent unit (GRU):
\begin{equation}
	\mathbf{S}^{(t+1)} = 
	\mathrm{GRU}\!\left(\mathbf{U}^{(t)},\mathbf{S}^{(t)}\right),
	\label{eq:gru_update}
\end{equation}
followed by a residual feed-forward network (FFN):
\begin{equation}
	\mathbf{S}^{(t+1)} = 
	\mathbf{S}^{(t+1)} + \mathrm{FFN}\!\left(\mathrm{Norm}(\mathbf{S}^{(t+1)})\right).
	\label{eq:sampling_ffn}
\end{equation}

This process is repeated for 3 iterations to get the final sampled feature. Note that the attention sampling module is used in both the encoder and decoder parts of the model, respectively performing downsampling and upsampling functions. During downsampling, the initial sampling point features $\mathbf{S}^{(0)}\in\mathbb{R}^{M\times D}$ are uniformly selected from the original point cloud features. For instance, when performing 4x downsampling, one point is selected every four points. For upsampling, the initial sampling point features are directly passed through the skip connection from the corresponding layer of the symmetric network.

\subsection{Parallel Flow-Conditioned Adaptive LayerNorm}
\label{sec:fca}
Different aerodynamic domains are characterized by distinct flow descriptors. For instance, for a car, usually only the traveling speed is needed to describe the environment. However, for an aircraft, in addition to the traveling speed (Mach number), the angle of attack (AoA) also needs to be introduced to describe the aircraft's climb or descent phases. To inject such domain-specific information within a unified backbone, UniField introduces Parallel Flow-Conditioned Adaptive Layer Normalization (PFC-AdaLN).

For a given domain $\textit{s}$, its flow condition vector $\mathbf{c}^{(s)}\in\mathbb{R}^{D_s}$ is mapped to per-channel modulation parameters via a specific MLP for domain s:
\begin{equation}
	(\boldsymbol{\mu}^{(s)},\boldsymbol{\sigma}^{(s)}) 
	= \mathrm{MLP}^{(s)}(\mathbf{c}^{(s)}),
	\label{eq:adaln_params}
\end{equation}
where $\boldsymbol{\mu}^{(s)},\boldsymbol{\sigma}^{(s)}\in\mathbb{R}^{D}$.

Given an intermediate feature tensor $\mathbf{X}\in\mathbb{R}^{N\times D}$, PFC-AdaLN first applies standard layer normalization without affine parameters:
\begin{equation}
	\hat{\mathbf{X}} = \mathrm{LN}(\mathbf{X}),
\end{equation}
and then modulates the normalized features as
\begin{equation}
	\mathbf{X}' = \hat{\mathbf{X}} \odot (1 + \boldsymbol{\sigma}^{(s)}) + \boldsymbol{\mu}^{(s)}.
	\label{eq:adaln_mod}
\end{equation}

To enable parallel computing, all domain-specific AdaLN branches are arranged in parallel, with a one-hot routing mechanism selecting the branch corresponding to domain $\textit{s}$. PFC-AdaLN is inserted before both the attention and feed-forward networks, enabling flow-aware modulation throughout the network.

\begin{figure*}[t!]
	\centering
	\includegraphics[width=2\columnwidth]{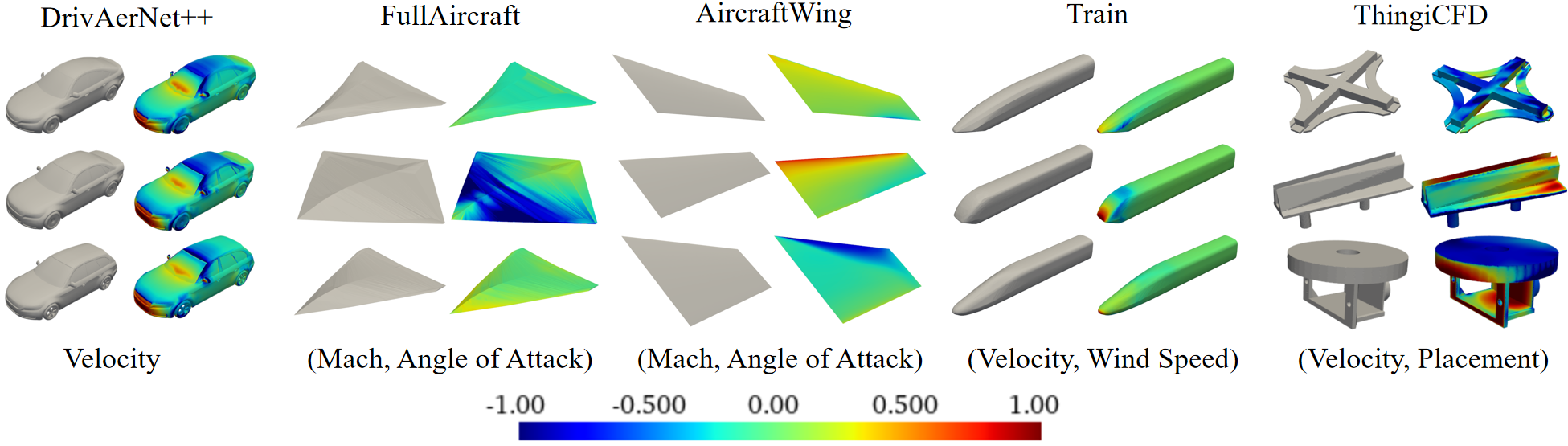}
	\caption{Visualization of the geometries and pressure distributions in each sub-datasets.}
	\label{fig: dataset}
	\vspace{-1em}
\end{figure*}

\begin{figure}[t!]
	\centering
	\includegraphics[width=1 \columnwidth]{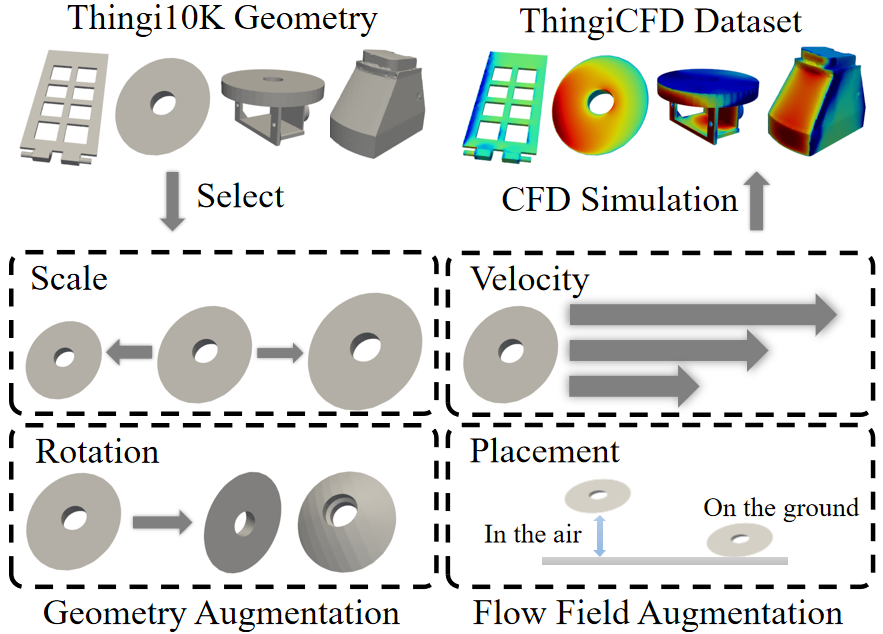}
	\caption{Process of generating ThingiCFD.}
	\label{fig: CFD}
	\vspace{-1em}
\end{figure}

\section{Datasets}

The scarcity of aerodynamic data arises from several factors: data generation requires large computational effort, and researchers typically focus only on data within their specialized domain. These issues make the existing data insufficient for training a model with strong cross-domain generalization capability. Therefore, we propose to integrate flow field data from multiple subfields to make up for the insufficiency of data in individual subfields. In addition, we also propose a novel general-shape CFD dataset, \textbf{ThingiCFD}, to further expand the distribution of the data. In Figure \ref{fig: dataset}, we provide visualizations of three samples for each dataset.

\paragraph{Data standardization}

To integrate data from different fields, we uniformly use the pressure coefficient as the prediction target, which is the dimensionless result of pressure, $C_p = \frac{P - P_0}{0.5 \rho V^2}$, where $P$ represents pressure, $P_0$ is the reference pressure such as atmospheric pressure, $\rho$ is the fluid density, and V is the velocity. In Figure \ref{fig: dataset}, we present three samples from each dataset, including geometry and pressure coefficients. During testing, for DrivAerNet++, to facilitate comparison with other models, we follow the official default settings: we use the original pressure value subtracting the mean (-94.5) and dividing by the standard deviation (117.25) as the ground truth for evaluation (and also perform the corresponding transformation on the model output results). For other datasets other than DrivAerNet++, we continue to use the pressure coefficient as the ground truth.

\paragraph{Integrated transportation dataset}

 We use four aerodynamic datasets spanning automobiles, trains, and aircraft. DrivAerNet++ \cite{DrivAerNet++} is a large-scale public automotive dataset, containing over 8,000 automobile samples with corresponding pressure fields under a fixed velocity of 30 m/s. The Train dataset \cite{qiu2025compressed} includes one high-speed train and two maglev train geometries, with surface pressure fields simulated under varying velocity and crosswind speed perpendicular to the traveling direction. As both land vehicles, automobiles and trains are combined in the first branch, where the flow condition is characterized by velocity and crosswind speed: $\mathbf{c^{(1)}=(v,w)}$. The Aircraft Wing and Full Aircraft datasets \cite{Wang_Zhang_Liu_Zhao_Lin_Chen_2025} cover 50 wing geometries and 100 full aircraft geometries, respectively. For each geometry, multiple samples are generated by varying the flow condition, which is characterized by the Mach number and Angle of Attack: $\mathbf{c^{(2)}=(Ma,AoA)}$.

\begin{table}[t]
	\begin{center}

		\setlength{\tabcolsep}{1mm}
		\begin{tabular}{l c c c c} 
			\toprule[2pt]
			\makebox[0.14\textwidth][c]{\multirow{3}{*}{\textbf{Model}}}  & 
			\multicolumn{4}{c}{\makebox[0.05\textwidth][c]{DrivAerNet++}} \\
			\cmidrule(r){2-5}
			& $\downarrow$MSE & $\downarrow$MAE & $\downarrow$RelL2 & $\downarrow$RelL1 \\
			& ($\times 10^{-2}$) & ($\times 10^{-1}$) & (\%) & (\%) \\
			\midrule
			RegDGCNN \shortcite{elrefaie2024drivaernet} 	& 8.29 & 1.61 & 27.72 & 26.21 \\
			Transolver \shortcite{Transolver} 	& 7.15 & 1.41 & 23.87 & 22.57 \\
			FigConvNet \shortcite{FIGConvNet} 	& 4.99 & 1.22 & 20.86 & 21.12 \\
			TripNet \shortcite{TripNet} 	& 5.14 & 1.25 & 20.05 & 20.93 \\
			\midrule
			\textbf{UniField} (ours) & \textbf{4.40} & \textbf{1.00} & \textbf{19.20} & \textbf{16.52} \\
			\bottomrule[2pt]
		\end{tabular}
		\caption{Model performance comparison on DrivAerNet++.}
		
		\label{tab: drivaernet++comparison}
	\end{center}
	\vspace{-1em}
\end{table}

\begin{table*}[t]
	\begin{center}
		\renewcommand\arraystretch{0.9}
		\setlength{\tabcolsep}{2.8mm}
		\begin{tabular}{c c c c c c c c} 
			\toprule
			\multirow{2}{*}{\textbf{Dataset}} & 
			\multicolumn{3}{c}{Training methods} &
			$\downarrow$MSE & $\downarrow$MAE  &
			$\downarrow$Rel L2 & $\downarrow$Rel L1 \\
			\cmidrule{2-4}
			&
			\textbf{Joint-Training}  & \textbf{+ ThingiCFD}  &
			\textbf{Single-Dataset} & ($\times 10^{-2}$) & ($\times 10^{-1}$) & (\%) & (\%) \\
			\midrule
			\multirow{3}{*}{\textbf{DrivAerNet++}} 
			& & & \checkmark & 4.47 & 1.04 & 19.61 & 16.88 \\
			& \checkmark & & & 4.50 & \textbf{1.00} & \textbf{19.10} & \textbf{16.40} \\
			& \checkmark & \checkmark & & \textbf{4.40} & \textbf{1.00} & 19.20 & 16.52 \\
			\midrule
			\multirow{3}{*}{\textbf{AircraftWing}} 
			& & & \checkmark & 2.09 & 0.56 & 26.5 & 19.1 \\
			& \checkmark & & & 1.42 & 0.47 & 22.7 & 15.7 \\
			& \checkmark & \checkmark & & \textbf{1.31} & \textbf{0.39} & \textbf{20.6} & \textbf{12.7} \\
			\midrule
			\multirow{3}{*}{\textbf{FullAircraft}} 
			& & & \checkmark & 0.40 & 0.29 & 27.8 & 19.9 \\
			& \checkmark & & & 0.22 & 0.24 & 20.5 & 16.2 \\
			& \checkmark & \checkmark & & \textbf{0.18} & \textbf{0.19} & \textbf{18.5} & \textbf{12.9} \\
			\midrule
			\multirow{3}{*}{\textbf{Train}} 
			& & & \checkmark & 2.18 & 0.45 & 146.9 & 106.5 \\
			& \checkmark & & & 1.06 & 0.35 & 110.0 & 85.7 \\
			& \checkmark & \checkmark & & \textbf{0.91} & \textbf{0.31} & \textbf{95.1} & \textbf{76.0} \\
			\bottomrule
		\end{tabular}
		\caption{
			Comparison between joint-training and single-dataset training. The inclusion or exclusion of ThingiCFD in joint training is indicated in the joint training is additionally listed.
			Joint training consistently improves prediction accuracy on datasets with limited samples, e.g., \textit{Train}, \textit{FullAircraft}, \textit{AircraftWing}, while large datasets such as \textit{DrivAerNet++} show comparable results under all the settings.
		}
		\label{tab: joint_vs_single}
		\vspace{-1em}
	\end{center}
	
\end{table*}

\paragraph{ThingiCFD}

Most existing aerodynamic pressure field datasets focus on specialized engineering domains with limited geometric variability, which restricts coverage of generic object-level flow scenarios. To address this limitation, we construct ThingiCFD, a large-scale CFD dataset based on the diverse geometries provided by Thingi10k \cite{Thingi10K}, a large-scale dataset of real-world 3D printing objects.

Figure \ref{fig: CFD} presents a construction diagram of ThingiCFD. Specifically, for each geometry from Thingi10k, we generate a corresponding CFD sample, resulting in a 10k-scale CFD dataset. By systematically randomizing both geometric and flow-related parameters, we substantially expand the data distribution across both geometric and flow-condition dimensions. The scale of objects is uniformly sampled within the range of 1–5 meters, and the object velocity is sampled from 0 to 100 m/s. To eliminate direction bias, each object is independently rotated around the three Cartesian axes by random angles uniformly sampled from 0 to 360 degrees. In addition, to account for different environmental interaction scenarios, each object is placed either on the ground or suspended in the air with equal probability, resulting in distinct boundary condition configurations.

Through this randomized simulation pipeline, ThingiCFD covers a wide spectrum of aerodynamic scenarios. All samples are represented by surface meshes with corresponding surface pressure coefficients obtained from CFD simulations. During training, ThingiCFD is used as an auxiliary data source jointly with other domain-specific datasets. In UniField, ThingiCFD samples are routed to the third branch, where the flow condition is described by the object’s velocity together with a binary indicator encoding whether the object is placed on the ground or in the air, i.e., $\mathbf{c^{(3)}=(v,P)}$.


\section{Experiments}

Based on our integrated dataset, we jointly train UniField on all  datasets and verify its performance via multiple experiments: 

1) We compare our model with existing models on the public automotive benchmark DrivAerNet++ and demonstrate that our model significantly reduces the prediction error of the surface pressure field compared with other models. 

2) To verify the performance advantage brought by joint training, we train UniField on each dataset separately with the same model scale and compare it with the jointly trained model. The results show that the performance of the jointly trained model is consistently better than that of the single-dataset-trained models, especially in fields with scarce data. For instance, for the \textit{train} and \textit{aircraft} datasets, the error of the jointly trained model is more than one-third lower than that of the single-dataset models. 

3) Considering that our method is mainly targeted at the aerodynamic subfield with scarce data, we investigated the impact of data volume on model performance. We conducted ablation experiments on the amount of data used for the aircraft wing dataset: We explore the scenario of training with less aircraft wing data while keeping the settings of other datasets unchanged. In this case, we observe that the jointly-trained model demonstrates a greater performance advantage.

\subsection{Benchmark Comparison on DrivAerNet++}

Following previous research \cite{FIGConvNet,TripNet}, we adopt the typical point cloud processing setting that randomly select 32,768 points from the original point clouds for training and testing. The comparison results are summarized in Table \ref{tab: drivaernet++comparison}. UniField consistently outperforms all other baseline methods, where UniField significantly reduces across all evaluation metrics, with MSE being more than 10\% lower than the previous best model FigConvNet. In terms of mean absolute error (MAE), UniField is 20\% lower than TripNet. Other metrics, including relative L1 and L2 losses (RelL1 and RelL2 in the table), are also significantly better than previous methods. These results establish UniField as achieving strong SOTA performance in surface pressure prediction.

\begin{figure*}[t!]
	\centering
	\includegraphics[width=2\columnwidth]{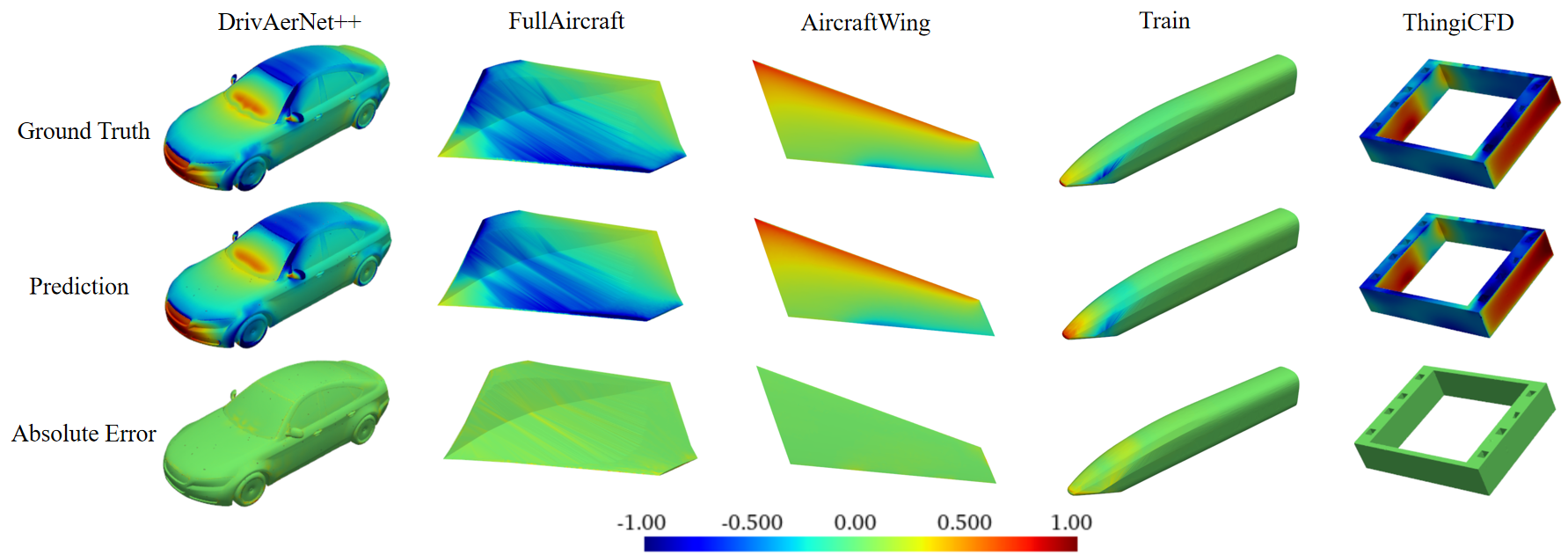}
	\caption{Surface pressure field prediction visualization. Each column corresponds to one dataset. The model successfully captures the main aerodynamic features across diverse domains, where the errors in most areas are close to zero. }
	\label{fig: visualization}
	\vspace{-1em}
\end{figure*}

\subsection{Joint-Training vs. Single Dataset Training}

\textbf{Setup:} To verify the benefits brought by the joint training strategy, we conducted a set of comparative experiments between the joint training models and the single-dataset models. Specifically, for each dataset, we trained a model from scratch using only that dataset and compared its performance with that of the joint training model on the same dataset. All other parameters are consistent except for the datasets used. All models are of the same scale. The jointly trained model is trained on the integrated dataset for 200 epochs, while the single-dataset models are trained for a similar number of steps as the jointly trained model.

\noindent \textbf{Results:} 
Table~\ref{tab: joint_vs_single} compares the performance of \textit{joint training} and \textit{single-dataset training} models across four aerodynamic datasets. A clear trend is that domains with limited data benefit more from joint training, whereas for large datasets, joint training shows marginal differences.

For instance, on the \textit{DrivAerNet++} dataset, which contains abundant automotive samples, UniField achieves marginal improvement with joint training. Furthermore, the \textit{Train} dataset, which are relatively small in scale, exhibit significant improvements when jointly trained with other domains: the MSE drops by more than half, from $2.18$ to $0.91$. Similarly, \textit{AircraftWing} and \textit{FullAircraft} also benefits notably from joint training (MSE $2.09$ to $1.31$ for \textit{AircraftWing} and $0.40$ to $0.18$ for \textit{FullAircraft}), indicating that the aerodynamic interactions are better modeled when auxiliary data from other geometries are introduced. These results verify that joint multi-domain training effectively mitigates data scarcity and improves generalization. It is also worth noting that within the framework of joint training, the inclusion of ThingiCFD has brought considerable performance improvements to data-scarce domains.

In addition, in Figure \ref{fig: visualization}, we present the predicted surface pressure field results of the jointly trained model on the test data of each subfield, showing that UniField can provide prediction results that are close to the ground truth, with errors in the vast majority of regions approaching zero. This suggests that UniField provides a viable and efficient data-driven complement to CFD for the considered aerodynamic scenarios.

\begin{table}[t]
	
	\begin{center}
		\small
		\setlength{\tabcolsep}{1.7mm}
		\begin{tabular}{c c c c} 
			\toprule[2pt]
			\multirow{2}{*}{\textbf{Data Volume}} &
			\multirow{2}{*}{\textbf{Joint-Training}} & $\downarrow$MSE & $\downarrow$MAE \\
			&& ($\times 10^{-2}$) & ($\times 10^{-1}$) \\
			\midrule
			\multirow{2}{*}{\textbf{1260 (90\%)}} 
			&            & 2.09 & 0.56 \\
			& \checkmark & \textbf{1.31} (\textcolor{customgreen}{$\downarrow$38\%}) & \textbf{0.39} (\textcolor{customgreen}{$\downarrow$30\%}) \\
			\midrule
			\multirow{2}{*}{\textbf{140 (10\%)}}	  
			&            & 4.33 & 1.06 \\
			& \checkmark & \textbf{1.91} (\textcolor{customgreen}{$\downarrow$56\%}) & \textbf{0.58} (\textcolor{customgreen}{$\downarrow$45\%}) \\
			
			\bottomrule[2pt]
		\end{tabular}
		\caption{
			Joint-training under data-scarcity situations.
			Results on the \textit{AircraftWing} dataset show that as the available training data decrease from 90\% to 10\% of the total, the performance gain of joint training becomes significantly larger.
			This highlights that joint multi-domain training is especially advantageous in data-scarce scenarios, where shared aerodynamic priors compensate for limited samples.
		}
		
		\label{tab: data_volume}
	\end{center}
	\vspace{-1em}
\end{table}

\subsection{Analysis of Joint Training and Data Scarcity}

\textbf{Setup:} One of the core purposes of our research is to propose flow field models that can be applied in areas where data is scarce. Therefore, we conducted a study on the volume of domain data. Specifically, for the aircraft wing dataset, we investigated the comparison between another set of single-dataset models and the joint training models, which are trained using only 140 aircraft wing samples, i.e., 10\% of the aircraft wing data, while the settings for the other datasets remained unchanged. We verify whether the joint training model can perform better with less data by comparing the gap between the single dataset model and the joint training model in the two scenarios.

\noindent \textbf{Results:} As shown in Table~\ref{tab: data_volume}, the advantage of joint training becomes more pronounced as the data volume decreases.
Under the full-data setting, the jointly trained model achieves about 30-40\% lower errors (MSE $1.32$ vs.\ $2.09$, MAE $0.39$ vs.\ $0.56$). 
However, when the available data are reduced to only 10\%, the performance gap widens substantially: the MSE of the joint-training model is merely one-third of that from single-dataset training ($1.91$ vs.\ $4.33$), and the MAE is also 45\% lower ($0.58$ vs.\ $1.06$). 
These results indicate that the more scarce the data is, the more performance gains can be provided by joint training.

\section{Conclusion}

In this work, we presented \textbf{UniField}, a universal framework for surface pressure field prediction across multiple aerodynamic domains, enabling joint multi-domain training to learn shared flow representations from diverse sources such as automobiles, trains, aircraft, and generic shapes.

Extensive experiments lead to three main conclusions. 1) UniField establishes new SOTA performance on the public DrivAerNet++ benchmark. 2) Joint training consistently outperforms single-dataset training, with particularly pronounced gains in data-scarce domains such as aircraft and train. 3) As the data in the target domain further decreases, joint training gradually demonstrates greater advantages.

Overall, UniField represents an important step toward foundation-style models for aerodynamic analysis, capable of leveraging heterogeneous datasets to learn robust and transferable surface field representations beyond single-domain learning.
Future work will extend UniField from surface pressure prediction to full flow field modeling, enabling large-scale, unified learning of complex aerodynamic phenomena.

\appendix

\bibliographystyle{named}
\bibliography{ijcai26}

\providecommand{\noopsort}[1]{}\providecommand{\singleletter}[1]{#1}%
\begin{thebibliography}{}

\bibitem[\protect\citeauthoryear{Chen \bgroup \em et al.\egroup
  }{2023}]{chen2023pointgpt}
Guangyan Chen, Meiling Wang, Yi~Yang, Kai Yu, Li~Yuan, and Yufeng Yue.
\newblock Pointgpt: Auto-regressively generative pre-training from point
  clouds.
\newblock {\em Advances in Neural Information Processing Systems},
  36:29667--29679, 2023.

\bibitem[\protect\citeauthoryear{Chen \bgroup \em et al.\egroup
  }{2025}]{TripNet}
Qian Chen, Mohamed Elrefaie, Angela Dai, and Faez Ahmed.
\newblock Tripnet: Learning large-scale high-fidelity 3d car aerodynamics with
  triplane networks, 2025.

\bibitem[\protect\citeauthoryear{Choy \bgroup \em et al.\egroup
  }{2025}]{FIGConvNet}
Chris Choy, Alexey Kamenev, Jean Kossaifi, Max Rietmann, Jan Kautz, and Kamyar
  Azizzadenesheli.
\newblock Factorized implicit global convolution for automotive computational
  fluid dynamics prediction, 2025.

\bibitem[\protect\citeauthoryear{Constantin and
  Foia{\c{s}}}{1988}]{constantin1988navier}
Peter Constantin and Ciprian Foia{\c{s}}.
\newblock {\em Navier-stokes equations}.
\newblock University of Chicago press, 1988.

\bibitem[\protect\citeauthoryear{Elrefaie \bgroup \em et al.\egroup
  }{2024}]{elrefaie2024drivaernet}
Mohamed Elrefaie, Angela Dai, and Faez Ahmed.
\newblock Drivaernet: A parametric car dataset for data-driven aerodynamic
  design and graph-based drag prediction.
\newblock In {\em International Design Engineering Technical Conferences and
  Computers and Information in Engineering Conference}, volume 88360, page
  V03AT03A019. American Society of Mechanical Engineers, 2024.

\bibitem[\protect\citeauthoryear{Elrefaie \bgroup \em et al.\egroup
  }{2025a}]{Elrefaie_2025}
Mohamed Elrefaie, Angela Dai, and Faez Ahmed.
\newblock Drivaernet: A parametric car dataset for data-driven aerodynamic
  design and prediction.
\newblock {\em Journal of Mechanical Design}, 147(4), March 2025.

\bibitem[\protect\citeauthoryear{Elrefaie \bgroup \em et al.\egroup
  }{2025b}]{DrivAerNet++}
Mohamed Elrefaie, Florin Morar, Angela Dai, and Faez Ahmed.
\newblock Drivaernet++: A large-scale multimodal car dataset with computational
  fluid dynamics simulations and deep learning benchmarks, 2025.

\bibitem[\protect\citeauthoryear{Gao \bgroup \em et al.\egroup
  }{2025}]{gao2025accurate}
Hongrui Gao, Tanghong Liu, Xiaodong Chen, Xiaoshuai Huo, Zhengwei Chen, Jie
  Zhang, and Boo~Cheong Khoo.
\newblock An accurate and efficient methodology on wind spectra relative to
  moving trains: field measurements of wind characteristics in complex
  terrains.
\newblock {\em Stochastic Environmental Research and Risk Assessment}, pages
  1--31, 2025.

\bibitem[\protect\citeauthoryear{Kovachki \bgroup \em et al.\egroup
  }{2023}]{kovachki2023neural}
Nikola Kovachki, Zongyi Li, Burigede Liu, Kamyar Azizzadenesheli, Kaushik
  Bhattacharya, Andrew Stuart, and Anima Anandkumar.
\newblock Neural operator: Learning maps between function spaces with
  applications to pdes.
\newblock {\em Journal of Machine Learning Research}, 24(89):1--97, 2023.

\bibitem[\protect\citeauthoryear{Li \bgroup \em et al.\egroup
  }{2020}]{li2020fourier}
Zongyi Li, Nikola Kovachki, Kamyar Azizzadenesheli, Burigede Liu, Kaushik
  Bhattacharya, Andrew Stuart, and Anima Anandkumar.
\newblock Fourier neural operator for parametric partial differential
  equations.
\newblock {\em arXiv preprint arXiv:2010.08895}, 2020.

\bibitem[\protect\citeauthoryear{Liu \bgroup \em et al.\egroup
  }{2021}]{liu2021swin}
Ze~Liu, Yutong Lin, Yue Cao, Han Hu, Yixuan Wei, Zheng Zhang, Stephen Lin, and
  Baining Guo.
\newblock Swin transformer: Hierarchical vision transformer using shifted
  windows.
\newblock In {\em Proceedings of the IEEE/CVF international conference on
  computer vision}, pages 10012--10022, 2021.

\bibitem[\protect\citeauthoryear{Luo \bgroup \em et al.\egroup
  }{2025}]{luo2025transolveraccurateneuralsolver}
Huakun Luo, Haixu Wu, Hang Zhou, Lanxiang Xing, Yichen Di, Jianmin Wang, and
  Mingsheng Long.
\newblock Transolver++: An accurate neural solver for pdes on million-scale
  geometries, 2025.

\bibitem[\protect\citeauthoryear{Qiu \bgroup \em et al.\egroup
  }{2025}]{qiu2025compressed}
Wei Qiu, Zhenxu Sun, Junxiang Wang, Ye~Bai, Shuanbao Yao, Dilong Guo, and
  Guowei Yang.
\newblock A compressed sensing based framework for surface pressure field
  reconstruction from sparse measurement.
\newblock {\em Physics of Fluids}, 37(4), 2025.

\bibitem[\protect\citeauthoryear{Raissi \bgroup \em et al.\egroup
  }{2019}]{RAISSI2019686}
M.~Raissi, P.~Perdikaris, and G.E. Karniadakis.
\newblock Physics-informed neural networks: A deep learning framework for
  solving forward and inverse problems involving nonlinear partial differential
  equations.
\newblock {\em Journal of Computational Physics}, 378:686--707, 2019.

\bibitem[\protect\citeauthoryear{Tran \bgroup \em et al.\egroup
  }{2021}]{tran2021factorized}
Alasdair Tran, Alexander Mathews, Lexing Xie, and Cheng~Soon Ong.
\newblock Factorized fourier neural operators.
\newblock {\em arXiv preprint arXiv:2111.13802}, 2021.

\bibitem[\protect\citeauthoryear{Wang and Wang}{2024}]{wang2024latent}
Tian Wang and Chuang Wang.
\newblock Latent neural operator for solving forward and inverse pde problems.
\newblock {\em Advances in Neural Information Processing Systems},
  37:33085--33107, 2024.

\bibitem[\protect\citeauthoryear{Wang \bgroup \em et al.\egroup
  }{2025}]{Wang_Zhang_Liu_Zhao_Lin_Chen_2025}
Yueqing Wang, Peng Zhang, Yushuang Liu, Jianing Zhao, Jie Lin, and Yi~Chen.
\newblock Aerodynamic coefficients prediction via cross-attention fusion and
  physical-informed training.
\newblock {\em Proceedings of the AAAI Conference on Artificial Intelligence},
  39(1):869--876, Apr. 2025.

\bibitem[\protect\citeauthoryear{Wu \bgroup \em et al.\egroup
  }{2024a}]{Transolver}
Haixu Wu, Huakun Luo, Haowen Wang, Jianmin Wang, and Mingsheng Long.
\newblock Transolver: A fast transformer solver for pdes on general geometries,
  2024.

\bibitem[\protect\citeauthoryear{Wu \bgroup \em et al.\egroup
  }{2024b}]{wu2024point}
Xiaoyang Wu, Li~Jiang, Peng-Shuai Wang, Zhijian Liu, Xihui Liu, Yu~Qiao, Wanli
  Ouyang, Tong He, and Hengshuang Zhao.
\newblock Point transformer v3: Simpler faster stronger.
\newblock In {\em Proceedings of the IEEE/CVF conference on computer vision and
  pattern recognition}, pages 4840--4851, 2024.

\bibitem[\protect\citeauthoryear{Yan \bgroup \em et al.\egroup
  }{2024}]{Yan_2024}
Chang Yan, Shengfeng Xu, Zhenxu Sun, Thorsten Lutz, Dilong Guo, and Guowei
  Yang.
\newblock A framework of data assimilation for wind flow fields by
  physics-informed neural networks.
\newblock {\em Applied Energy}, 371:123719, October 2024.

\bibitem[\protect\citeauthoryear{Ye \bgroup \em et al.\egroup
  }{2025}]{ye2025pdeformer}
Zhanhong Ye, Zining Liu, Bingyang Wu, Hongjie Jiang, Leheng Chen, Minyan Zhang,
  Xiang Huang, Qinghe~Meng Zou, Hongsheng Liu, Bin Dong, et~al.
\newblock Pdeformer-2: A versatile foundation model for two-dimensional partial
  differential equations.
\newblock {\em arXiv preprint arXiv:2507.15409}, 2025.

\bibitem[\protect\citeauthoryear{Zehtabiyan-Rezaie \bgroup \em et al.\egroup
  }{2022}]{10.1063/5.0091980}
Navid Zehtabiyan-Rezaie, Alexandros Iosifidis, and Mahdi Abkar.
\newblock Data-driven fluid mechanics of wind farms: A review.
\newblock {\em Journal of Renewable and Sustainable Energy}, 14(3):032703, 06
  2022.

\bibitem[\protect\citeauthoryear{Zhang \bgroup \em et al.\egroup
  }{2024}]{zhang2024prediction}
Qiao Zhang, Xuan Zhao, Kai Li, Xinwu Tang, Jifei Wu, and Weiwei Zhang.
\newblock Prediction model of aircraft hinge moment: Compressed sensing based
  on proper orthogonal decomposition.
\newblock {\em Physics of Fluids}, 36(7), 2024.

\bibitem[\protect\citeauthoryear{Zhang \bgroup \em et al.\egroup
  }{2025}]{ZHANG2025100571}
Hao Zhang, Yang Shen, Wei Huang, Zan Xie, and Yao-Bin Niu.
\newblock Deep transfer learning for three-dimensional aerodynamic pressure
  prediction under data scarcity.
\newblock {\em Theoretical and Applied Mechanics Letters}, 15(2):100571, 2025.

\bibitem[\protect\citeauthoryear{Zhou and Jacobson}{2016}]{Thingi10K}
Qingnan Zhou and Alec Jacobson.
\newblock Thingi10k: A dataset of 10,000 3d-printing models.
\newblock {\em arXiv preprint arXiv:1605.04797}, 2016.

\bibitem[\protect\citeauthoryear{Zhou \bgroup \em et al.\egroup
  }{2024}]{zhou2024unisolver}
Hang Zhou, Yuezhou Ma, Haixu Wu, Haowen Wang, and Mingsheng Long.
\newblock Unisolver: Pde-conditional transformers are universal pde solvers.
\newblock {\em arXiv preprint arXiv:2405.17527}, 2024.

\bibitem[\protect\citeauthoryear{Zou \bgroup \em et al.\egroup
  }{2025}]{zou2025joint}
Junhong Zou, Wei Qiu, Zhenxu Sun, Xiaomei Zhang, Zhaoxiang Zhang, Xiangyu Zhu,
  and Zhen Lei.
\newblock Joint optimization of sensor placement and sparse pressure field
  reconstruction with a two-stage framework for limited data.
\newblock {\em Physics of Fluids}, 37(7), 2025.

\end{thebibliography}

\end{document}